\pdfoutput=1
\documentclass[12pt]{article} 
\usepackage{amsmath}
\usepackage{graphicx,psfrag}
\usepackage{enumerate}
\usepackage{hyperref} 

\usepackage{natbib}

\newcommand{\blind}{0}

\begin{document}

\bibliographystyle{natbib} 

\def\spacingset#1{\renewcommand{\baselinestretch}%
{#1}\small\normalsize} \spacingset{1}


\if0\blind
{
  \title{\bf Top-Frequency Parallel Coordinates Plots}
  \author{Vincent Yang \\
    University of California, Davis \\
    and \\
    Harrison Nguyen \\
    University of California, Davis \\
    and \\
    Norman Matloff \\
    University of California, Davis \\
    and \\
    Yingkang Xie \\
    Northwestern University  
    }
  \maketitle
} \fi


\bigskip
\begin{abstract}

Parallel coordinates plotting is one of the most popular methods for
multivariate visualization. However, when applied to larger data sets,
there tends to be a ``black screen problem,'' with the screen becoming
so cluttered and full that patterns are difficult or impossible to
discern.  Xie and Matloff (2014) proposed remedying this problem by
plotting only the most frequently-appearing patterns, with frequency
defined in terms of nonparametrically estimated multivariate density.
This approach displays ``typical'' patterns, which may reveal important
insights for the data.  However, this remedy does not cover variables
that are discrete or categorical.  An alternate method, still
frequency-based, is presented here for such cases.  We discretize all
continuous variables, retaining the discrete/categorical ones, and plot
the patterns having the highest counts in the dataset.  In addition, we
propose some novel approaches to handling missing values in parallel
coordinates settings.

\end{abstract}

\noindent%
\vfill

\newpage
\spacingset{1.45} 
\section{Introduction}
\label{sec:intro}

The problem of visualizing data of more than three dimensions has vexed
analysts throughout the history of ``number crunching,'' but has become
especially acute in today's era of Big Data. For the common case of $n$
data points, each consisting of $p$ variables/features, Big Data has
either $n$ or $p$ large, often both.  As noted in \cite{kane}, in
analyses of Big Data it is useful to distinguish between the large-$p$
and large-$n$ cases.  Though discussions of visualizing multidimensional
data typically focus on the large-$p$ setting, here we will see that
large $n$ can be problematic as well.  

One of the most popular approaches has been {\it parallel coordinates
plots} (PCPs); see e.g.\ \cite{inselberg} and \cite{theus}.  Here one
draws $p$ vertical axes, with each data point being represented by a
polygonal line connecting nodes on the axes. Denoting data point $i$ by
$(X_{i1},...,X_{ip})$, the node at axis $j$ has height $X_{ij}$.  

This is simple and satisfying for small values of $n$ and $p$.  However,
in general it can be a challenge to tease insight out of these
fascinating lines.  For instance, for $p$ larger than, say, two dozen,
there simply is not enough room on a computer screen to view all of the
axes at once.  Indeed, even if there is sufficent room, it may be
difficult to view the relationship of variables that are far apart on
the screen. Software packages generally handle this in different ways,
allowing the user to permute the columns or even generate a new
permutation every few seconds.

\begin{figure}
\begin{center}
\includegraphics{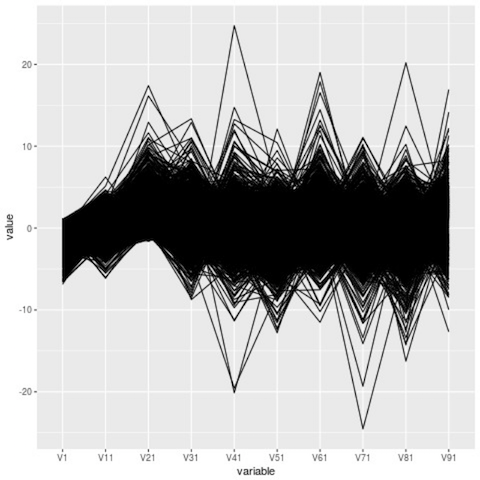} 
\end{center}
\caption{Million song subset, ordinary parallel coordinates
\label{fig:millblack}}
\end{figure}

With large $n$ there typically is a ``black screen problem'' (BSP), in
which substantial portions of the screen become solid black.  For
instance, consider the Million Song Dataset in \cite{thierry}. One of the
available versions has $n = 515345$ and $p = 91$.  The first column of
the data is the year in which the song was released, and the remaining
data involves technical audio aspects of the song.  Even using just a
random subsample of 50000 data points, and just every 10th column, we
encounter the black screen problem shown in Figure \ref{fig:millblack}
(the R package {\bf ggparcoord} was used here.)

\section{Top-Frequency Parallel Coordinates Plots: Continuous-Variables Case}

There have been numerous methods proposed for handling the BSP.  One
approach for instance is to make the printed points more transparent,
via {\it alpha blending}, as in \cite{theus}.  But this merely postpones
the problem, as $n$ grows.  

Another approach is to draw a random subsample of size $N$ from the
data, consisting of say, hundreds or thousands of data points, and then
form a PCP from the subsample.  This ``thinning out'' of the data
ameliorates the BSP, but even a graph of a few hundred lines may become
crowded and difficult to interpret, and there are other issues. We will
return to this in Section \ref{subsample}.

By contrast, our approach ``thins out'' the data in a different way, by
using the entire data set but plotting only the most representative
points.  (For investigating outliers, this becomes plotting the {\it
least} representative points.)

To see how this works, first consider settings in which the variables
are continuous, as opposed to discrete or categorical.  In
\cite{xiematloff} the third and fourth authors of the present paper
proposed plotting only those points having the highest estimated
multivariate density, and implemented the method in the CRAN package
{\bf freqparcoord}, \cite{freqparcoord}.  These then are ``typical''
points, thus hopefully shedding some light on the multivariate
relations.  

In a sense, this is analogous to cluster hunting, say finding (an
unknown number of) components of a Gaussian mixture.  Typically in
cluster analysis the centroid (vector of means) or medoid (vector of
univariate medians) is taken as the anchor of a component; here it is
the multivariate mode of the component.

Let us refer to this as Top-Frequency Parallel Coordinates (TFPC).
Though we will often be discussing specific software packages here, this
paper will treat TFPC as a general method.  Throughout this paper, we
will use the letter $F$ to denote the number of lines plotted, i.e.\ we
plot the $F$ most-frequent lines.

In {\bf freqparcoord} the density estimation is performed using
k-Nearest Neighbors estimation.  To estimate the density of a
$q$-dimensional random vector $W$ at a point $t$, based on a sample
$W_1,...,W_m$, we first determine

\begin{equation}
D(t) = \max_{W_i \textrm{ in } A(t)} \textrm{distance}(W_i,t)
\end{equation}

\noindent
where $A(t)$ is the set of the $k$ closest sample points to $t$

The density estimate is then

\begin{equation}
\frac{k}{\textrm{vol}(D(t))}  
\end{equation} 

\noindent
with vol() representing the volume of a hypersphere with radius given in
the argument.  Since we are simply comparing distances, we can ignore
a multiplicative constant in that volume, and simply take the volume to
be $\pi^q$.  The lines plotted are then the ones with the $k$ highest
estimated frequencies.

Again, the crucial point is the `T' in ``TFPC.''  We only plot the most
frequently-occurring points, as determined via the density estimation.
This is key to avoiding the BSP.  By contrast, \cite{wegman} for example
proposed plotting entire density estimates, resulting in a shaded graph. 

\begin{figure}
\begin{center}
\includegraphics{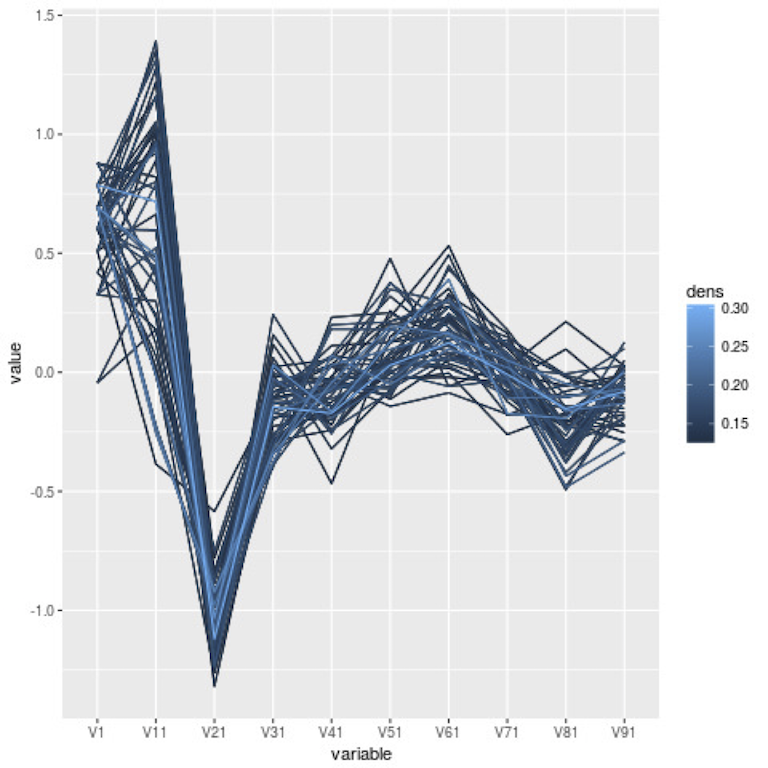} 
\end{center}
\caption{Million song subset, freqparcoord
\label{fig:millfreqparcoord}}
\end{figure}

A typical example of TFPC, applied in this case to the song data ($n =
50000$, plotting the $F = 50$ most frequent lines), is shown in Figure
\ref{fig:millfreqparcoord}.  There is a narrow but solid band running
through the middle of the jumble of lines.  It is lighter gray in
monochrome, lighter blue in the color version.  Tracing the band back to
the first variable, Year, one finds that the band corresponds
approximately to years 2004 onward (scaled value about 0.6).  This is a
good example of what PCPs are supposed to do, culling out unusual
patterns that may call the analyst's attention to interesting underlying
phenomena.  What is happening in this case?

As noted, the band is light and narrow.  The lightness of the band ---
higher estimated density --- merely signifies that songs in the dataset
are disproportionately from later years.  What is key, though, is the
narrowness, which tells us that for the years 2004 onward, there was
much less variation within each variable V2, V3 and so on than among the
previous years.  The use of Auto-Tune made things more uniform.

So, what happened around that time period?  In 1997, a technology
debuted called Auto-Tune, which performs corrections on a singer's
voice.  Within a few years, virtually all song recordings were using
Auto-Tune, resulting in the reduced variation seen in the figure.
(The effect will be more clearly brought out with another
technique, to be introduced below.)

\begin{figure}
\begin{center}
\includegraphics{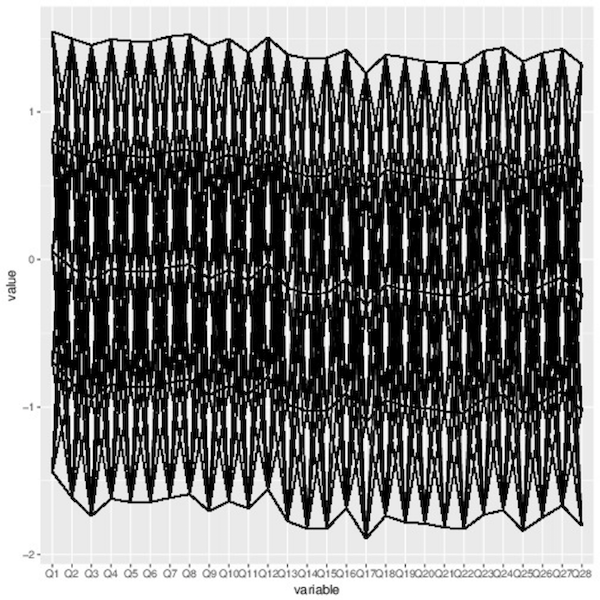} 
\end{center}
\caption{Turkish evaluations, ordinary parallel coordinates
\label{fig:turkggpar}}
\end{figure}

Here is another example, using another well-known dataset, Turkish
Student Instructor Evaluations, from the UCI Machine Learning Data
Repository, \cite{lichman}.  Here $n = 5820$ and $p = 28$.  (Five
categorical variables have not been included.)  The 28 ratings, on a 1-5
scale, included criteria such as ``The course aims and objectives were
clearly stated at the beginning of the period.'' The result, plotted
using standard parallel coordinates, is shown in Figure
\ref{fig:turkggpar}.
This is not quite black-screen, but certainly lacking in any discernible
pattern.

\begin{figure}
\begin{center}
\includegraphics{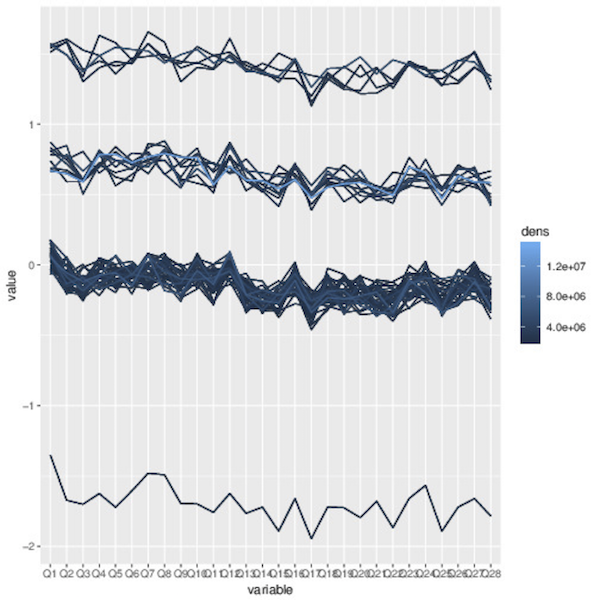} 
\end{center}  
\caption{Turkish data with jitter, freqparcoord  
\label{fig:turkfreqparcoord}}
\end{figure}

This dataset would at first be considered an ideal candidate for using
TFPC, but this actually presents a problem, since the data are discrete
rather than continuous. In fact, the k-NN estimation process encounters
a divide-by-0 problem, as a neighborhood often has volume 0.  After
applying R's {\bf jitter()} function (default value) and running {\bf
freqparcoord}, again with $F = 50$, we have Figure
\ref{fig:turkfreqparcoord}.

Here the lines are rather flat, with little variation, even with the
jitter.  The implication is remarkable:  Typically, a student gives
identical responses to all 28 questions!  Apparently the surveyors need
not ask so many questions after all.

\section{Discrete/Categorical Data}

Most real datasets contain at least one categorical variable, typically
many.  TFPC methodology, aimed at illustrating continuous variables, does
not work directly.  Distance measures no longer make sense, as the data
is non-ordinal.  

\begin{figure}
\begin{center}
\includegraphics{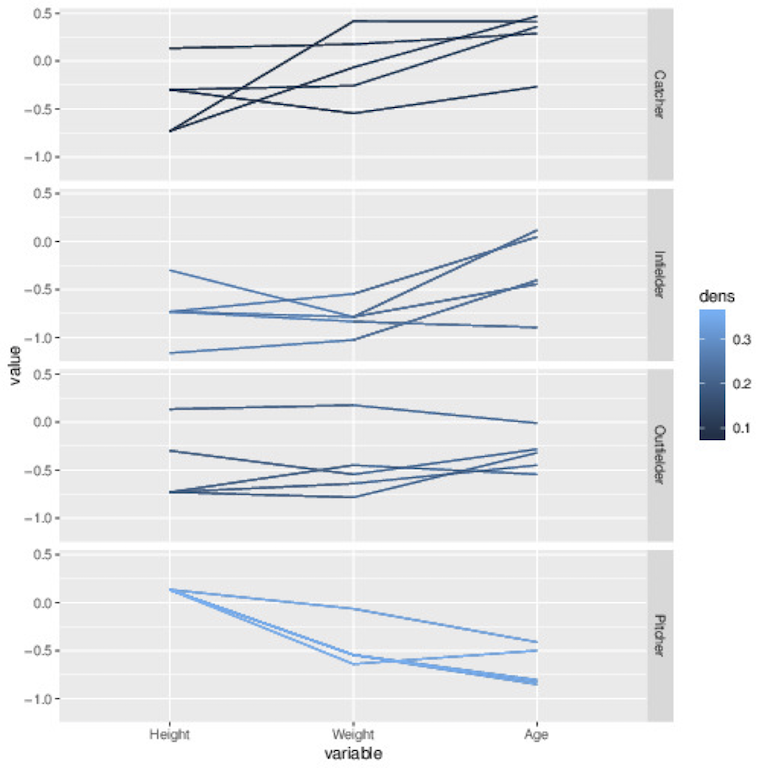} 
\end{center}
\caption{MLB data, freqparcoord
\label{fig:mlb1}}
\end{figure}

Our {\bf freqparcoord} package handles this by stacking separate plots
atop each other.  To illustrate this, consider the dataset {\bf mlb} on
U.S.\ major league baseball players, included in the
package.\footnote{Data provided courtesy of the UCLA Statistics Dept.}
Figure \ref{fig:mlb1} shows the result for $F = 5$, plotting height, weight
and age, for positions Catcher, Infielder, Outfielder and Pitcher.

Again, the data is centered and scaled, with those parameters being
determined by the data as a whole, not within-group.  Look in particular
at the weight variable:  The catchers seem to be on the heavy side,
confirming the popular image of ``beefy'' catchers.

However, this approach is not feasible if a categorical variable has
many levels.  For example, in the {\bf mlb} data, there are 30 teams,
and certainly no room on a screen to stack 30 graphs.  And this would be
compounded when using the player position variable in addition to team,
necessitating $30 \times 4 = 120$ graphs.  Converting to dummy variables
would not be a solution either, with no room on the screen for the
additional 119 columns.

The solution is to abandon using multivariate density as our measure of
frequency.  Instead, we make all variables discrete (including
discretizing the continuous ones), and then simply measure frequency of
a pattern by its count in the dataset.  

This discrete/categorical case is the main focus of the present paper.
It is implemented in our later package, {\bf cdparcoord},
\cite{cdparcoord}.

\begin{figure}
\begin{center}
\includegraphics{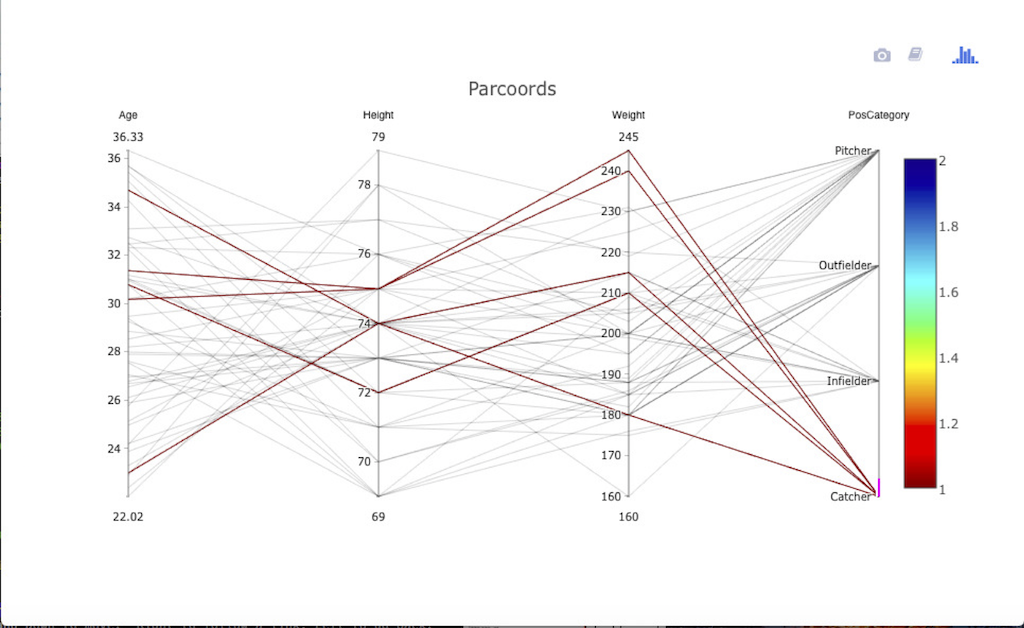} 
\end{center}
\caption{MLB data, cdparcoord
\label{fig:mlb2}}
\end{figure}

As our first example, consider the {\bf mlb} data.  We applied a {\bf
discretize()} function from the package, which (in default
configuration) converts the numeric variables to {\bf nlevels}
equally-spaced quantiles.  We then plotted the $F = 50$ most common
lines, and also performed a couple of mouse operations.  The first mouse
action allows one to drag a column to a new location, which we used to
move the age variable to the far left, so that height and weight would
be adjacent to player position.  The second mouse operation involved
{\it brushing}, implemented as clicking and dragging the Catcher node in
the player position column up a bit, thus highlighting the catchers.
(That portion of the Catcher axis then becomes highlighted in magenta,
to indicate where the brushing was applied.) The result is shown in
Figure \ref{fig:mlb2}.  Note that the colored legend on the right shows
frequencies.

Here we see that the catchers not only tend to be somewhat heavier than
average, but also somewhat shorter.  The combination of those two
traits, a ``stocky'' build, is again consistent with a general
perception regarding catchers.

\begin{figure}
\begin{center}
\includegraphics{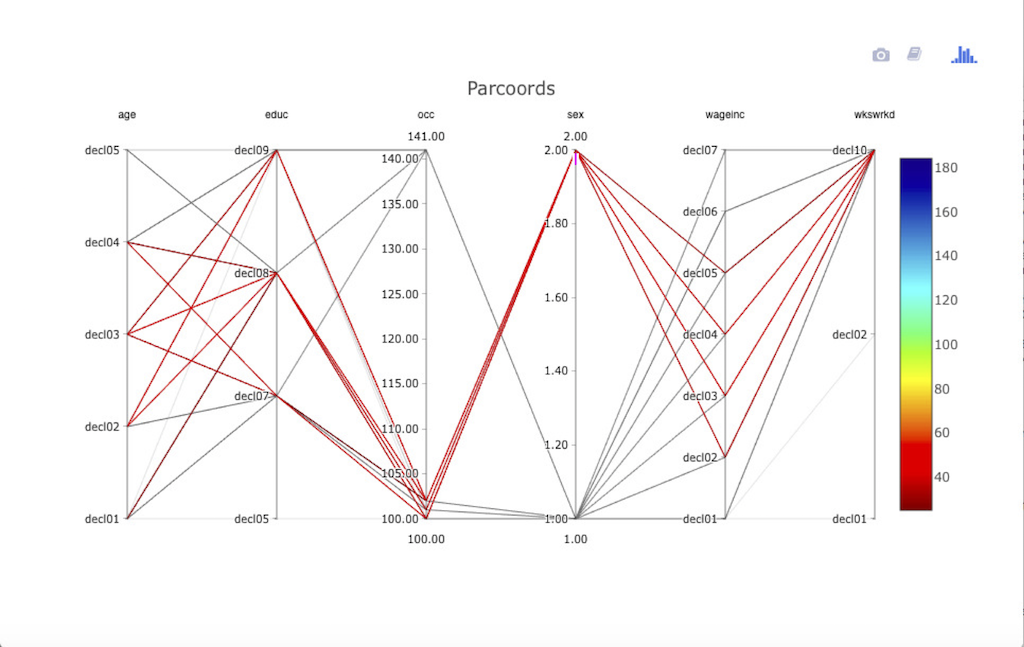} 
\end{center}
\caption{Programmer and engineer  data, cdparcoord
\label{fig:prgeng}}
\end{figure}

Another dataset included in the package is {\bf prgeng}, involving
programmers and engineers in the Silicon Valley area in the 2000 U.S.\
census.  We removed those with salaries above \$250,000, and plotted the
top 150 lines, again brushing the female lines.  The result is shown in
Figure \ref{fig:prgeng}.  Are the women underpaid?

It does appear that the women (sex code 2) tend to have lower wages
(shown in the column {\bf wageinc}).  They do work in occupations that
are somewhat lesser paid (100, 101, {\bf occ} column) with Programmer
titles, not Software Engineer (thus more likely in a bank, say, than a
tech company), and not with Hardware Engineer titles (code 141).  On the
other hand, the males with the same job codes still seem to tend to have
higher wages than their female peers.  The important question of gender
discrimination in Silicon Valley cannot be solved with such preliminary
analysis and limited number of variables, but the example does show how
TFPC may be used.

\begin{figure}
\begin{center}
\includegraphics{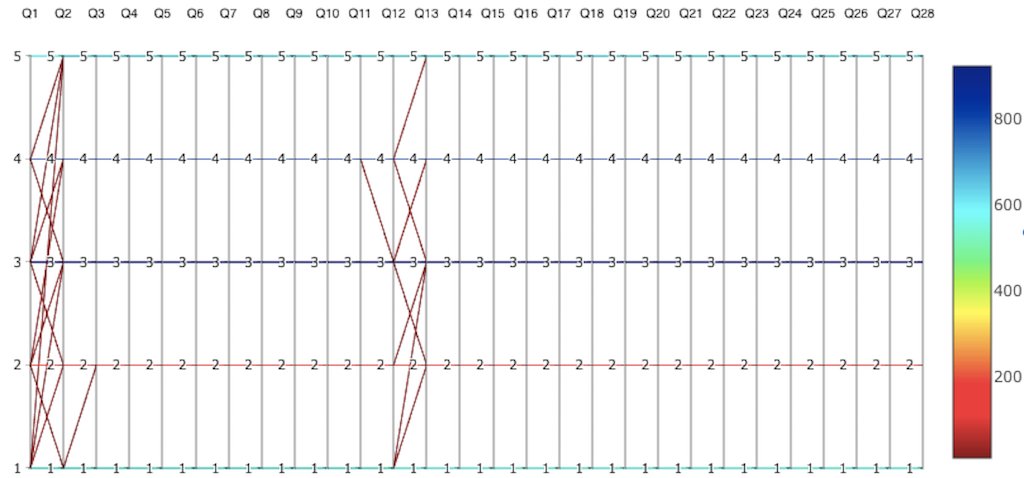} 
\end{center}
\caption{Turkish evaluation data, cdparcoord
\label{fig:turkfpc}}
\end{figure}

Returning to the Turkish student evaluation data, with {\bf cdparcoord},
with discrete / categorical TFPC there is now no need to add jitter, and
with 25 lines we have Figure \ref{fig:turkfpc}.  Again, the most common
lines are seen to be horizontal, meaning that students tend to give
identical answers to the various questions. This plot does seem to
indicate some nonconstancy within certain questions --- e.g.\ an
interesting drop from a 4 rating to a 2 from Question 11 to 13 --- but
again it appears that many more questions are being asked than is
necessary.

\begin{figure}
\begin{center}
\includegraphics{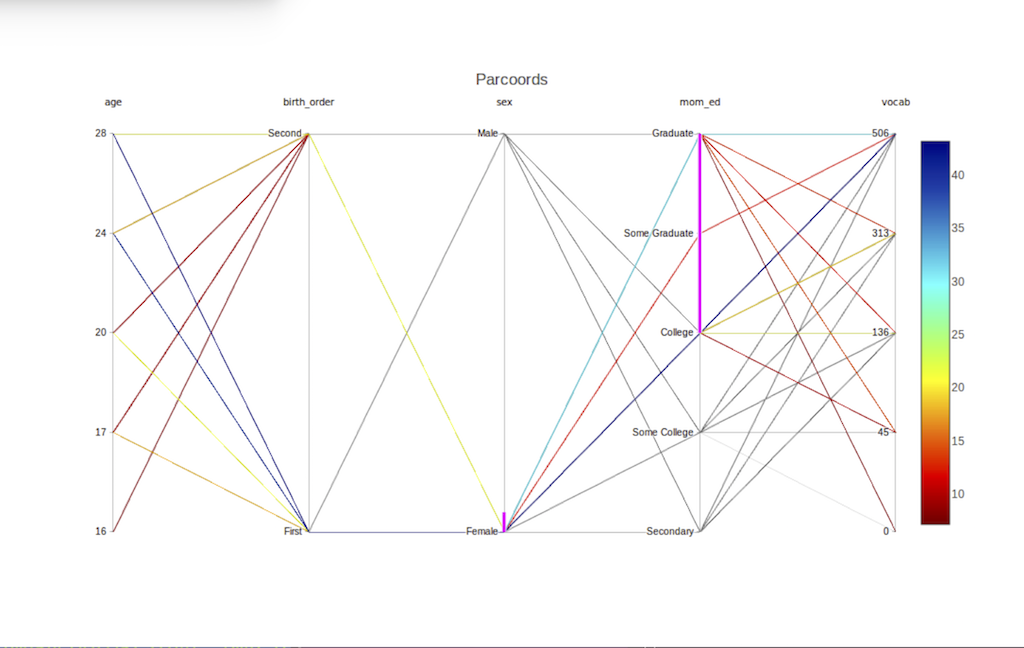} 
\end{center}
\caption{Stanford WordBank data, cdparcoord
\label{fig:wb}}
\end{figure}

It is important to be able to condition on more than one variable/value
at a time.  Consider for instance the Stanford WordBank vocabulary data,
\cite{wordbank}, involving very young children.  The plot is shown in
Figure \ref{fig:wb}.  Here we have focused on girls of mothers having at
least a college education, indicated by the magenta segments of two of
the axes. Note that, at least at these very young ages, the relation
between the mother's education and the daughter's vocabulary size is not
as strong as one might guess {\it a priori}.

\section{Outlier Hunting}

One advantage of frequency-based PCPs is that one
can plot the {\it least}-frequent lines in order to search for possible
outliers.  Both the {\bf freqparcoord} and {\bf cdparcoord}
packages allow for this, signified by the user specifying a negative
number of lines to be plotted.

\begin{figure}
\begin{center}
\includegraphics{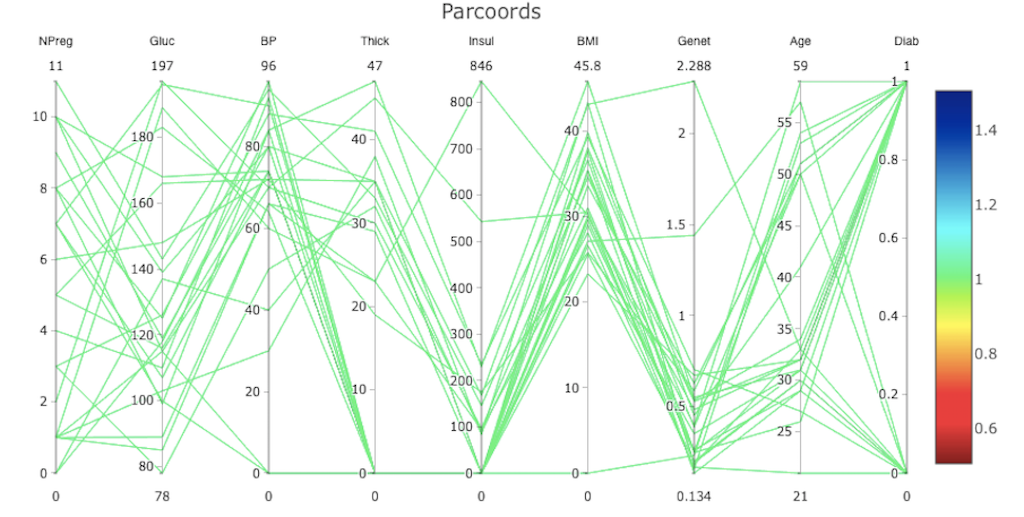} 
\end{center}
\caption{Diabetes data, outliers
\label{fig:pimaoutliers}}
\end{figure}

Here is an example, using the famous Pima diabetes dataset from the UCI
repository.  The call

\begin{verbatim}
discparcoord(pima,k=-25)
\end{verbatim}

\noindent yields Figure \ref{fig:pimaoutliers}. This is rather
startling, as it shows that there are people in the dataset with 0
values for BP, BMI and insulin, a medical impossibility.  Note too that
at least one observation has more than one 0 (there are several), a fact
we learn due to PCPs being {\it multivariate} methods.  These
observations should probably be removed from the data set.

\begin{figure}
\begin{center}
\includegraphics{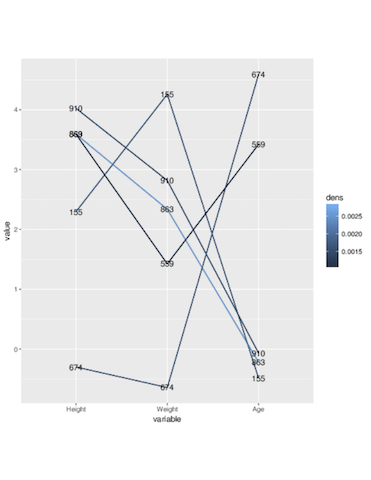} 
\end{center}
\caption{MLB data, outliers
\label{fig:mlboutliers}}
\end{figure}

The {\bf freqparcoord} package allows one to optionally label lines with
their index numbers within the dataset, making it easier to identify the
outlier cases for possible action. (In {\bf cdparcoord}, one has the
option of saving a text file with all cases and their frequencies.) Here
is an example using the {\bf mlb} data, specifying 5 lines, in Figure
\ref{fig:mlboutliers}.

Case 674 looks interesting, with extremely high age but somewhat low
height and weight.  Here is the record in the dataset:

\begin{verbatim}
        Name Team      Position Height Weight   Age  PosCategory
Julio_Franco  NYM First_Baseman     73    188 48.52  Infielder
\end{verbatim}

\noindent
Further investigation reveals that Franco later was still playing professional
baseball at age 57.

\section{Cluster Hunting}

As noted earlier, TFPC may be viewed as similar to cluster analysis.
However, since TFPC finds only the most frequent tuples, we run the risk
of missing what amount to rare clusters.

In our {\bf freqparcoord} package, this is handled via a {\bf locmax}
option.  As the name implies, it specifies to display a tuple not
according to whether it is among the overall highest frequencies but
according to whether it has the highest frequency in a neighborhood.

In {\bf cdparcoord} this approach is infeasible, due to the lack of a
metric.  However, one can force a known rare category to display by
giving it extra weight in the tuple counting, via an {\bf accentuate}
argument.

\section{Handling Missing Values}

A problem with many real data sets is the prevalence of missing values,
coded NA in R.  Normally this would not be an issue with exploratory
graphical methods such as parallel coordinates, which do not rely on
precise point estimation.  However, TFPC requires complete cases,
and for large $p$ the chance of a particular case being complete may be
quite low.  This would result in a much smaller dataset to plot.  

Take the Stanford WordBank vocabulary data mentioned earlier.  There
only 2741 of the 5498 cases are complete, with only $p = 12$ variables.
Datasets having a much larger $p$ could be even worse in this regard.  

Here we propose several remedies.  There is a vast literature on formal
methods for handling missing values [\cite{little}].  Here we use only
the most restrictive of the major assumptions, Missing Completely at
Random (MCAR), and then the somewhat less restrictive Missing at Random
(MAR). We also propose a heuristic method.

It will be convenient to use conditional probability mass function
notation $\pi$.  For any (possibly incomplete) vector $X'$ in our
sample. let $X = (U,V)$ denote the true values of the components,  where
$U$ and $V$ denote the observed and unobserved components of the tuple;
for simplicity we take $V$ to consist of a single component here.  Let
$M$ denote an indicator variable for the event that $V$ is missing.
MCAR states that 

\begin{equation}
\pi_{M ~|~ U,V} = \pi_M
\end{equation}

\noindent
i.e.\ missingness is entirely independent of the data.

MAR is a little more subtle:

\begin{equation}
\label{mar}
\pi_{M ~|~ U,V} = \pi_{M ~|~ U} 
\end{equation}

\noindent
i.e.\ missingness may be affected by the observed components but not the
unobserved one.  A commonly-offered example of MAR involves a mental
health survey, and possible nonresponse. Let $U$ denote gender.  Suppose
male subjects are generally more reluctant to respond to a question 
on depression than are females, but suppose such reluctance in either 
gender is not affected by the degree of depression ($V$).  Then MAR
would hold.


In the following sections, we present outlines of three possible
approaches.

\subsection{Method of Moments Estimation}

Here we assume MCAR.  Let $X$ denote a randomly generated tuple from the
population, including possibly unobserved components, and let $X'$
denote the observed tuple, including NAs. 

Let $\mathcal{T}$ denote the set of all possible tuples. Say for example
we have $p = 6$ variables, each taking on the values 1, 2 or 3.
$\mathcal{T}$ then would be the 6-fold Cartesian product of the set
$\{1,2,3\}$.  For any $t \in \mathcal{T}$, let $p_t$ denote the
population proportion for that tuple, i.e.\ $P(X = t)$.

Now let $\mathcal{T}'$ denote the superset of $\mathcal{T}$ that
includes tuples with NA values.  In the above example, (2,2,1,3,NA,3)
would be a member of this set, as would (NA,2,1,3,NA,3), and as would
all of $\mathcal{T}$.  For any $t' \in \mathcal{T}'$, let $M(t') \subset
\mathcal{T}$ be the collection of all tuples $t  \in \mathcal{T}$ that
become $t'$ upon suitable replacement of components by NA.  In ther
words, $M(t')$ is the set of intact tuples from which $t'$ may have
come.  For any $t'$, we will use $c(t')$ to denote the count of NAs in
$t'$.

Finally, let $q$ denote the probability of an NA value. Under MCAR and
the typical additional assumption that missingness is independent across
components of a tuple, the probability that an observed tuple has NAs in
$m$ specified components is $(1-q)^{p-m}q^m$.

The natural estimate of $q$ is the proportion of NAs in the $np$ tuple
components in our data.  We will take that estimate to be the true
population value, but an alternating procedure along the lines of the EM
algorithm could be used.

Then for any $t' \in \mathcal{T}'$, we have 

\begin{equation}
P(X' = t') = 
(1-q)^{p-c(t')} q^{c(t')}
\sum_{t \in M(t')} p_t 
\end{equation}

Replace the left-hand side here by its direct sample estimate, the
proportion of observed tuples that equal $t'$, and continue to assume
that $q$ is known.  As we vary $t'$, this gives us a system of linear
equations in the $p_t$.\footnote{Note that some of these equations will
be redundant, and that we must use the fact that the $p_t$ sum to 1.}
Solving for the $p_t$ yields estimates for those quantities,
$\widehat{p}_t$.

We can then determine the $F$ largest values of the $\widehat{p}_t$, and 
generate a TFPC plot that thus takes into account both the intact
and partially-missing data.

\subsection{An Update Method}

Here we assume MAR, again for the sake of notational simplicity taking
$V$ to consist of a single component.  Missingness for that component
will be denoted by $M$.

Using (\ref{mar}) we start with

\begin{eqnarray}
\pi_{V | M,U} &=& 
\pi_{V,M,U} ~/~ \pi_{M,U} \\
&=& \pi_{M | U,V} \cdot \pi_{U,V} ~/~ \pi_{M,U} \\
\label{mar1}
&=& \pi_{M | U} \cdot \pi_{U,V} ~/~ \pi_{M,U} 
\end{eqnarray}

All three expressions in the final equation can be estimated from our
data:  $\pi_{U,V}$ can be estimated from our complete-case data, while
$\pi_{M | U}$ and $\pi_{M,U}$ can be estimated from our partially-observed
data.

As a simple example, suppose $p = 2$ and our data is

\bigskip

\begin{tabular}{|r r|}
\hline
U & V \\ \hline
1 & 2 \\
3 & 2 \\
3 & NA \\
3 & 2 \\
3 & 1 \\
2 & 2 \\
\hline
\end{tabular}

\bigskip

Our tuple frequency table based on the intact cases is

\bigskip

\begin{tabular}{|r r|r|}
\hline
U & V & Freq \\ \hline
1 & 2 & 1 \\
3 & 2 & 2 \\
3 & 1 & 1 \\
2 & 2 & 1 \\
\hline
\end{tabular}

\bigskip

Then we have the following estimates:

\begin{eqnarray}
\widehat{P}(M = 1 ~|~ U = 3) = 1/4 \\ 
\widehat{P}(U = 3, V = 1) = 1/5 \\ 
\widehat{P}(M = 1, U = 3) = 1/6 
\end{eqnarray}

\noindent
From (\ref{mar1}), we have 

\begin{equation}
\widehat{P}(V = 1 ~|~ M = 1, U = 3) =  (1/4) \cdot (1/5) ~/~ (1/6) =
3/10
\end{equation}

\noindent
Similarly, $\widehat{P}(V = 2 ~|~ M = 1, U = 3) = 2/5$.
and thus   $\widehat{P}(V = 1 ~|~ M = 1, U = 3) = 1/10$.

We can thus ``update'' our tuple frequency table, counting the (3,NA)
tuple 3/10 for (3,1), 2/5 for (3,2) and 1/10 for (3,3): 

\bigskip

\begin{tabular}{|r r|r|}
\hline
U & V & Freq \\ \hline
1 & 2 & 1 \\
3 & 2 & 2.6 \\
3 & 1 & 1.3 \\
2 & 2 & 1 \\
3 & 3 & 0.1 \\
\hline
\end{tabular}

\bigskip

In that manner, we can proceed iteratively, with one update for each
non-intact observation.  At any stage, we could use the new frequency
table in calculating the quantities in (\ref{mar}).  

The final frequency table would then be used in forming the parallel
coordinates plot.

\subsection{A Heuristic Approach}

The MCAR and MAR models are of course restrictive.  For instance, MCAR treats
missingness as being independent across variables, a condition that may
not hold in some data sets.

In the Stanford WordBank data mentioned earlier, for instance, about
13.5\% of the data values are NAs.  Under MCAR, the probability of a row
being complete would then be $(1 - 0.135)^{13}$, about 0.152.  Yet the
actual proportion is $2741/5498$, about 0.499.  In other words, the NA
values tend to clump.

Again, the missing-values field is vast, and the methods are
assumption-laden.  Furthermore, one must keep in mind the implications
of the fact that the cases having missing values may be systemically
different from the others.  Nevertheless, a simple heuristic may be of
value.  One is included in our package {\bf cdparcoord}, as a
tuning parameter {\bf NAexp}, which we now describe.\footnote{As of this
writing, the Method of Moments and Update approaches are not implemented
in our package.}

For concreteness, consider this simple example, with $p = 6$ and with
each variable taking on the integers between 1 and 3.  Suppose we have
the observation (2,2,1,3,NA,3).  It could have been (2,2,1,3,1,3),
(2,2,1,3,2,3) or (2,2,1,3,3,3).  Intuitively, it would seem wrong to
ignore the information that this observation mostly matches these three
tuples.

One way to use such information would be to give $(1/3) \times (5/6) =
5/18$ credit for each of the three possibilities.  In our package {\bf
cdparcoord}, there is a tuning parameter {\bf NAexp} that allows
the user to choose how much to count such partial matches.  In the above
example, the 5/6 figure would be taken to the {\bf NAexp} power.

\section{Why Not Just Subsample?}
\label{subsample}

As noted earlier, one approach to the BSP is to draw a random subsample
of size $N$ from the data, consisting of say, hundreds or thousands of
data points, and then form a PCP from the subsample.  However, the loss
of information due to subsampling may obscure important trends or create
spurious ones especially if $p/N$ is large, essentially the multiple
inference problem.  Important outliers may also be missed.

To compare the subsampling and TFPC approaches, consider the diamonds
data, bundled with the {\bf ggplot2} package.  We took a subsample of
size 2500 (out of almost 54000), and ran an ordinary PCP on the
subsample.\footnote{This was the function {\bf parallelplot} from the
{\bf lattice} package.} We then ran {\bf cdparcoord} on the the full
data set, but with the $F = 2500$ most-frequent lines.  

\begin{figure}
\begin{center}
\includegraphics{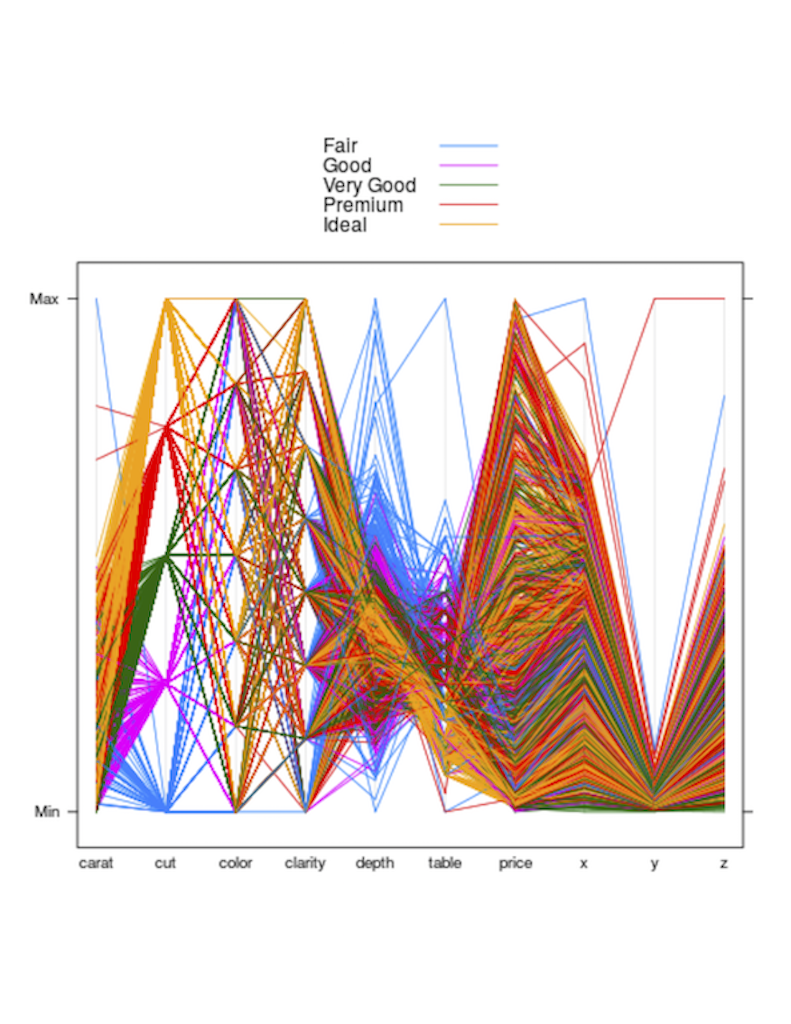} 
\end{center}
\caption{Diamond data, PCP on subsample, $N = 2500$ 
\label{fig:dilat}}
\end{figure}

\begin{figure}
\begin{center}
\includegraphics{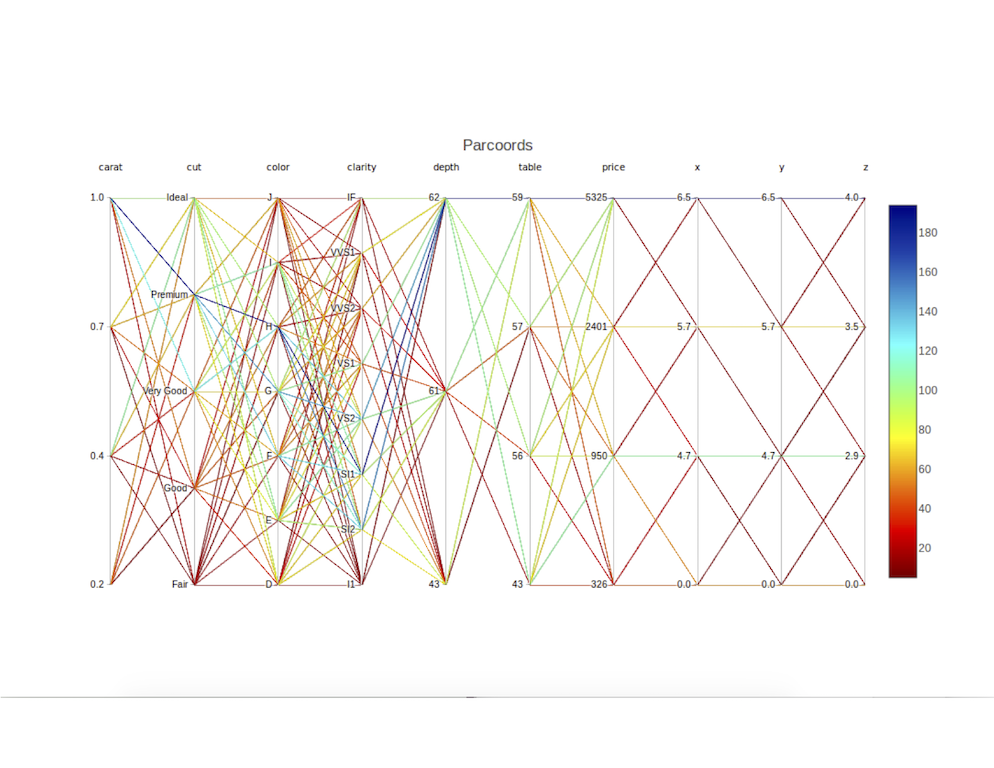} 
\end{center}
\caption{Diamond data, {\bf cdparcoord}, $F = 2500$
\label{fig:dicd}}
\end{figure}

The results are shown in Figures \ref{fig:dilat} and
\ref{fig:dicd}.\footnote{Due to discretization, there is much
overplotting, giving an appearance of many fewer to 2500 lines.}
Looking at the Premium grade, for instance, there is a clear trend in
the TFPC version, seen in the blue line.  This is much harder to see in
the subsampled, standard PCP.

\section{Computational Issues}

Though the tabulation of tuple frequencies could easily be parallelized,
there would seem to be no need even for fairly large datasets.  For
instance, simulated data with $n = 1000000$ and $p = 150$ and 10 values
for each variable took only about 20 seconds to process on a low-end
laptop.  In any event, parallel computation of frequency counts is
offered in {\bf cdparcoord}.

Implementation of the NA models proposed above could be more intensive
in both computational time and memory space.  Parallelization of the
solution of linear systems brings only limited speedup
[\cite{matloff2015parallel}], and again, memory space may be an issue.

\section{Discussion and Conclusions}
\label{sec:conc}

There has been much work on visualization methods for the categorical
data case, such as \cite{unwin} and \cite{ggparallel}.  All of them have
settings in which they work well. In this paper, we add TFPC as another
possible choice for those interested in data visualization.

For advanced usage, there is elegant geometric theory that can aid in
acquiring insight from the data [\cite{inselberg}].  For example, say
two variables that are adjacent in a parallel coordinates plot have a
strong negative correlation.  Then the lines between these two columns
will tend to approximately converge at a common point between the
columns.  However, there is no insight to be gained if one has the black
screen problem, and TFPC is one solution to the latter.

As pointed out in \cite{murrell}, techniques for categorical data tend
to be lesser known than continuous-variable methods, even though a
number of useful methods have been developed.  There are for instance as
{\it parallel sets}, \cite{kosara} and {\it mosaic plots}, \cite{theus}.
However, though these are highly effective for a small number of
variables, there typically is not enough room on the screen, or in the
viewer's perceptive abilities, to display larger numbers of variables.

Thus we believe parallel coordinates methods, including TFPC, holds
great promise. But even putting the ``black screen problem'' aside,
users cannot expect to necessarily have quick, automatic ``Eureka!''
responses when viewing such a plot.  These issues were described well in
Ben Shneiderman's Forward to \cite{inselberg}. While he speaks of ``the
cleverness and power'' of the parallel coordinates approach, Shneiderman
admits to ``struggling'' with interpreting some of these plots.  

The benefits of the parallel coordinates approach do require patience.
In addition, as an exploratory tool, the analyst may find it
useful to generate several plots of the same data, changing for example
the number of lines plotted, the order of the columns, the use of
brushing and so on.  

But as the examples presented here show, the TFPC approach can
yield excellent insight, including in datasets having many categorical
variables.  The ``black screen problem'' is handled in a simple, natural
manner.

\bigskip
\begin{center}
{\large\bf SUPPLEMENTARY MATERIAL}
\end{center}

{\bf Code listings:}

For Figure \ref{fig:millblack}:

\begin{verbatim}
# had earlier generated and saved a 50K subsample from
# https://archive.ics.uci.edu/ml/datasets/YearPredictionMSD
library(data.table)
ms <-
   fread('~/DataSets/MillionSong/FiftyKSongs.csv',header=TRUE)
ms <- as.data.frame(ms)
ms <- ms[,seq(1,91,10)]
ggparcoord(ms,1:10)
\end{verbatim}

For Figure \ref{fig:millfreqparcoord}:

\begin{verbatim}
library(cdparcoord)  # brings in freqparcoord auto
# ms from above
freqparcoord(ms,m=50)
\end{verbatim}

For Figure \ref{fig:turkggpar}:

\begin{verbatim}
turk <- 
   read.csv('~/DataSets/TurkEvals/turkiye-student-evaluation.csv',   
      header=TRUE)
turk <- turk[,-(1:5)]
ggparcoord(turk,m=50)
\end{verbatim}

For Figure \ref{fig:turkfreqparcoord}:

\begin{verbatim}
# turk as above
turkj <- turk
for (i in 1:28) turkj[,i] <- jitter(turk[,i])
freqparcoord(turkj,m=50)
\end{verbatim}

For Figure \ref{fig:mlb1}:

\begin{verbatim}
data(mlb)
freqparcoord(mlb,5,4:6,7) 
\end{verbatim}

For Figure \ref{fig:mlb2}:

\begin{verbatim}
# mlb as above
discparcoord(mlb[,4:7],k=50)
# click and drag Age to far left
# click Catcher, see +, then drag up a bit
\end{verbatim}

For Figure \ref{fig:prgeng}:

\begin{verbatim}
data(prgeng)
pe <- prgeng[,c(1,3,5,7:9)] 
pe1 <- pe 
pe25 <- pe1[pe1$wageinc < 250000,] 
pe25 <- makeFactor(pe25,c('educ','occ','sex'))
pe25disc <- discretize(pe25,nlevels=5)  
discparcoord(pe25disc,k=150) 
\end{verbatim}

For Figure \ref{fig:turkfpc}:

\begin{verbatim}
# turk as above (but no jitter added)
trk <- turk
# convert ints to factors, so have e.g. 2 not 2.00
for (i in 1:28) trk[,i] <- as.factor(turk[,i])
discparcoord(trk,k=25)
\end{verbatim}

For Figure \ref{fig:wb}:

\begin{verbatim}
wb <- wb[,c(2,5,7,8,10)] 
wb <- discretize(wb,nlevels=5) 
wb <- reOrder(wb,'mom_ed',
   c('Secondary','Some College','College','Some Graduate','Graduate'))
discparcoord(wb,k=100)
\end{verbatim}

For Figure \ref{fig:pimaoutliers}:

\begin{verbatim}
pima <-
   read.csv('~/Research/DataSets/Pima/pima-indians-diabetes.data',
      header=TRUE) 
discparcoord(pima,k=-25)
\end{verbatim}

For Figure \ref{fig:mlboutliers}:

\begin{verbatim}
# mlb as above
freqparcoord(mlb,-5,4:6,plotidxs=TRUE)
\end{verbatim}

For Figure \ref{fig:dilat}:

\begin{verbatim}
library(lattice)
ds <- diamonds[sample(nrow(diamonds), 2500),]
parallelplot(~ds, group = cut, data = ds, horizontal.axis = FALSE,
             auto.key = TRUE)
\end{verbatim}

For Figure \ref{fig:dicd}:  

\begin{verbatim}
dd <- discretize(diamonds,nlevels=4) 
discparcoord(dd,k=2500)
\end{verbatim}


 
\bibliography{Paper}

\end{document}